\documentclass[twocolumn,showpacs,superscriptaddress,amssymb,10pt]{revtex4-1}

\usepackage{graphicx}
\usepackage{dcolumn}
\usepackage{bm}
\usepackage{epsfig}
\usepackage{color}
\usepackage{longtable}

\definecolor{aogreen}{rgb}{0.0, 0.5, 0.0}

\def\ketm#1{  \left\vert  #1   \right\rangle   }

\def\bram#1{  \left\langle  #1   \right\vert   }

\def\sprm#1#2{  \left\langle #1 \left\vert \right. #2 \right\rangle   }

\def\mem#1#2#3{  \left\langle #1 \left\vert  #2 \right\vert #3 \right\rangle   }

\def\redmem#1#2#3{  \left\langle #1 \left\Vert
                  #2 \right\Vert #3 \right\rangle   }
\def\twobytwo#1#2#3#4{  \left( \begin{array}{cc}
                                   #1 & #2   \\[0.2cm]
                                   #3 & #4   \end{array} \right)   }

\definecolor{mymainmessagecolor}{RGB}{10,200,10}
\begin{document}

\preprint{}
\title{
Relativistic calculations of the non-resonant two-photon ionization of neutral atoms
}

\author{J.~Hofbrucker}
\affiliation{Theoretisch-Physikalisches Institut, Friedrich-Schiller-Universit\"at Jena, Max-Wien-Platz 1, D-07743
Jena, Germany}%
\affiliation{Helmholtz-Institut Jena, Fr\"o{}belstieg 3, D-07743 Jena, Germany}

\author{A.~V.~Volotka}
\affiliation{Helmholtz-Institut Jena, Fr\"o{}belstieg 3, D-07743 Jena, Germany}%
\affiliation{Department of Physics, St.~Petersburg State University, Oulianovskaya 1, 198504 St.~Petersburg, Russia}%

\author{S.~Fritzsche}
\affiliation{Theoretisch-Physikalisches Institut, Friedrich-Schiller-Universit\"at Jena, Max-Wien-Platz 1, D-07743
Jena, Germany}
\affiliation{Helmholtz-Institut Jena, Fr\"o{}belstieg 3, D-07743 Jena, Germany}%


\begin{abstract}

The non-resonant two-photon one-electron ionization of neutral atoms is studied theoretically in the framework of relativistic second-order perturbation theory and independent particle approximation. In particular, the importance of relativistic and screening effects in the total two-photon ionization cross section is investigated. Detailed computations have been carried out for the $K$-shell ionization of neutral Ne, Ge, Xe, and U atoms. The relativistic effects significantly decrease the total cross section, for the case of U, for example, they reduce the total cross section by a factor of two. Moreover, we have found that the account for the screening effects of the remaining electrons leads to occurrence of an unexpected minimum in the total cross section at the total photon energies equal to the ionization threshold, for the case of Ne, for example, the cross section drops there by a factor of three.

\end{abstract}

\newpage

\maketitle

\section{Introduction}
\label{Sec.Introduction}

Two-photon one-electron ionization is one of the fundamental non-linear processes in the light-matter interaction with various spectroscopic applications. In the past, however, most experiments were focused on the two-photon ionization of atomic outer shell electrons \cite{Duncanson/PRL:1976, Siegel/JPB:1983, Dodhy/PRL:1985, Wang/PRA:2000}. Only with the recent advancements in free-electron lasers (FEL), the study of non-linear processes in atoms and molecules at extreme ultraviolet and x-ray energies became feasible \cite{Pellegrini/RMP:2016} and renewed an interest in inner-shell excitation and ionization processes.  
One of the first experiments utilising FEL facilities used electron or ion spectrometers to study the two-photon ionization of 4$d$ electron of neutral Xe atom \cite{Richardson/PRL:2010} and 1$s$ electron of heliumlike Ne$^{8+}$ ion \cite{Doumy/PRL:2011}. 
Meanwhile, modern FEL facilities reach beam intensities of about $I \approx 10^{20}$ W/cm$^2$, although the non-resonant two-photon ionization of inner-shell electrons still remains a challenge due to small cross sections. Experimentally, the two-photon ionization can be measured by collecting the $K$-fluorescence. This fluoresce radiation emitted by bound electrons decaying into the $K$-shell vacancy is a direct signature of the two-photon ionization process.
This experimental approach has been utilized in the case of $K$-shell ionization of neutral Ge \cite{Tamasaku/NP:2014} and Zr \cite{Ghimire} atoms.

First detailed calculations of the two-photon ionization of atomic hydrogen were performed by Zernik within non-relativistic dipole approximation \cite{Zernik/PR:1964} more than 50 years ago. In this work, he also introduced the well-known $Z^{-6}$ scaling of the total cross section with the nuclear charge $Z$ for non-resonant two-photon ionization and, hence, provided an estimate of a total cross section for all hydrogenlike ions. However, later in Refs. \cite{Koval/JPB:2003, Koval/JPB:2004, Koval/Dissertation}, complete relativistic calculations were carried out which demonstrated that quite strong deviation from this scaling occurs due to relativistic effects. Recently, in Refs. \cite{Florescu/PRA:2011, Florescu/PRA:2012}, the relativistic effects have also been investigated in the two-photon above-threshold ionization of low-$Z$ hydrogenlike ions. Although the relativistic effects have been found rather important, no systematic relativistic calculations have been carried out until now for the two-photon ionization of neutral atoms.

In this paper, we investigate the relativistic and screening effect contributions to the total two-photon $K$-shell ionization cross section of neutral atoms. In particular, we consider photon energies below the ionization threshold so that no single photon $K$-shell ionization is possible. In Sec. \ref{Sec.Theory}, we first formulate the relativistic second-order perturbation theory, based on the Dirac equation for describing the non-resonant two-photon ionization. By using, in addition, the independent particle approximation and particle-hole formalism, we are able to reduce the many-electron transition amplitude to an effective \textit{single-electron} amplitude, from which an expression for the total two-photon ionization cross section is obtained. In section \ref{Sec.Computation}, we then outline the numerical procedure that is employed in this work. Detailed calculations are carried out for the non-resonant two-photon $K$-shell ionization of neutral Ne, Ge, Xe, and U atoms. The total cross section as a function of energy is compared for hydrogenlike and neutral systems in Sec. \ref{Sec.ResultsAndDiscussion}. In this section, we also demonstrate that the relativistic effects need to be taken into account in the two-photon ionization cross section calculations of neutral atoms, especially for high-$Z$ atoms. Moreover, we show that although screening effects tend to increase the cross section, they result in an unexpected minimum of the total cross section in the near-threshold ionization energy region. Finally, a summary is given in Sec. \ref{Sec.SummaryAndOutlook}.

Relativistic units ($\hbar=c=m=1$) are used throughout the paper, unless stated otherwise.

\section{Theoretical background}
\label{Sec.Theory}

We here consider the two-photon one-electron ionization of neutral atoms. This process can be expressed as follows

\begin{eqnarray}
\ketm{\alpha_i J_i M_i}+\gamma_1+\gamma_2 \rightarrow \ketm{\alpha_f J_f M_f}+\ketm{\bm{p}_e m_e},
\end{eqnarray}
where the atom is initially in the many-electron state $\ketm{\alpha_i J_i M_i}$, with total angular momentum $J_i$, its projection $M_i$, and where $\alpha_i$ denotes all further quantum numbers that are needed for unique characterization of the state. After the simultaneous interaction of the atom with two photons $\gamma_1$ and $\gamma_2$ with energies $\omega_1$ and $\omega_2$, respectively, the system is in a final state $\ketm{\alpha_f J_f M_f, \bm{p}_e m_e}$. The system now consists of a singly charged ion $\ketm{\alpha_f J_f M_f}$ with a hole in the substate $\ketm{a}$, as well as a continuum electron $\ketm{\bm{p}_e m_e}$, with well-defined asymptotic momentum $\bm{p}_e$ and spin projection $m_e$. 
In the following subsection, we will use the particle-hole formalism and the independent particle approximation in order to reduce the many-electron transition amplitude to a one-electron transition amplitude. Then, employing the density matrix formalism, we derive an expression for the total non-resonant two-photon ionization cross section. \vspace{0.1cm}

\subsection{Evaluation of transition amplitude}
\label{Subsec.TransitionAmplitude}

In second-order perturbation theory, the transition amplitude for the two-photon single-electron photoionization of an atom in the initial state $\ketm{\alpha_iJ_iM_i}$ into a final state $\ketm{\alpha_f J_f M_f, \bm{p}_e m_e}$ under the simultaneous absorption of two photons with wave vectors $\bm{k}_1,\bm{k}_2$ and polarization vectors $\bm{\hat{\varepsilon}}_{\lambda_1}, \bm{\hat{\varepsilon}}_{\lambda_2}$ can be written as 

\begin{widetext}

\begin{eqnarray}
\label{GeneralTransitionAmplitude}
M_{J_i M_i J_f M_f m_e}^{\lambda_1 \lambda_2}&=&~\int\kern-1.5em\sum_{\nu}\frac{
	\mem{\alpha_fJ_fM_f,\bm{p}_e m_e}
		{\hat{\mathcal{R}}(\bm{k}_2,\bm{\hat{\varepsilon}}_{\lambda_2})}
		{\alpha_{\nu}J_{\nu}M_{\nu}}
	\mem{\alpha_{\nu}J_{\nu}M_{\nu}}
		{\hat{\mathcal{R}}(\bm{k}_1,\bm{\hat{\varepsilon}}_{\lambda_1})}
		{\alpha_iJ_iM_i}		
		}
	{E_{i}+\omega_1-E_{\nu}}\\ \nonumber
	&&\hspace{4cm}+(\bm{k}_1\leftrightarrow \bm{k}_2,\bm{\hat{\varepsilon}}_{\lambda_1}\leftrightarrow \bm{\hat{\varepsilon}}_{\lambda_2},\omega_1 \leftrightarrow \omega_2 ).
\end{eqnarray}
\end{widetext}
For the general case of two inequivalent photons, the additional term  $(\bm{k}_1\leftrightarrow \bm{k}_2,\bm{\hat{\varepsilon}}_{\lambda_1}\leftrightarrow \bm{\hat{\varepsilon}}_{\lambda_2},\omega_1 \leftrightarrow \omega_2 )$ arises from the interchange of the interaction sequence of the two photons with the atom. The evaluation of expression (\ref{GeneralTransitionAmplitude}) requires a summation to be carried out over the complete spectrum of intermediate states $\ketm{\alpha_{\nu}J_{\nu}M_{\nu}}$. The operator $\hat{\mathcal{R}}$ denotes the one-particle transition operator describing the electron-photon interaction. This operator can be represented in the second quantization formalism (see, e.g., \cite{Johnson/Book}) as 

\begin{eqnarray}
\label{OperatorExpansion}
\hat{\mathcal{R}}(\bm{k},\bm{\hat{\varepsilon}}_{\lambda})=
	\sum_{lm}
	\mem{l}{\alpha_{\mu} A^{\mu}_{\lambda}(\omega)}{m}
	a^{\dag}_{l} a_{m},
\end{eqnarray}
where $\ketm{m}, \ketm{l}$ are the single-electron initial and final states, $a^{\dag}_l$ and $a_m$ are the corresponding electron creation and annihilation operators, $\alpha_{\mu}$ denotes the four-vector of the Dirac matrices and $A^{\mu}_{\lambda}=({\phi_{\lambda},\bm{A}_\lambda})$ is the photon wavefunction.

Due to the interaction of the atom with the two photons, an electron from a substate $\ketm{a}\equiv\ketm{n_a j_a l_a m_a}$ of the atom is promoted into a continuum state, leaving a hole (or vacancy) in the atomic subshell. Here, $n_a$ is the principal quantum number, $l_a$ is the orbital angular momentum, $j_a$ and $m_a$ are the total angular momentum and its projection, respectively. According to the particle-hole formalism, a state with a hole in a substate $\ketm{n_a j_a l_a m_a}$ has angular momentum properties of a particle with angular momentum $j_a$ and its projection $-m_a$. Then, within the independent particle approximation, the final state after an ionization process is obtained by applying the electron creation ($a^{\dag}_{p_e m_e}$) and annihilation ($a_{n_a j_a l_a m_a}$) operators to the initial state and coupling the initial atom and hole angular momenta using a Clebsh-Gordan coefficient $\sprm{.. ..}{..}$. 
Hence, the final state of the system can be expressed as

\begin{eqnarray}
\label{FinalMultiElectronWaveFunction}
\ketm{\alpha_f J_f M_f, \bm{p}_e m_e}&=&
	\sum_{m_a M} \sprm{j_a,-m_a,J_i, M}{J_f M_f} \\ \nonumber 
	&\times &
	(-1)^{j_a-m_a}a^{\dag}_{p_e m_e}a_{n_a j_a l_a m_a}\ketm{\alpha_i J_i M}.
\end{eqnarray}
If we therefore insert expressions (\ref{OperatorExpansion}) and (\ref{FinalMultiElectronWaveFunction}) into Eq. (\ref{GeneralTransitionAmplitude}), apply the electron creation and annihilation operators and carry out the summation over the magnetic quantum number $M$, the many-electron transition amplitude reduces to an amplitude which only depends on one-electron wavefunctions of the active electron

\begin{eqnarray}
\label{SimplifiedTransitionAmplitude}
&M&^{\lambda_1 \lambda_2}_{J_i M_i J_f M_f m_e}\\\nonumber
	&=&\sum_{m_a}(-1)^{j_a-m_a} \nonumber
	\sprm{j_a, -m_a, J_i, M_i}{J_f, M_f} \\\nonumber
	&\times& \int\kern-1.5em\sum_{~n}
	\frac{
		\mem{\mathbf{p}_e m_e}
			{\alpha_{\mu} A^{\mu}_{\lambda_2}(\omega_2)}
			{n}
		\mem{n}
			{\alpha_{\mu} A^{\mu}_{\lambda_1}(\omega_1)}
			{a}		
			}
		{E_{n_a j_a}+\omega_1-E_{n_n j_n}}\\
		&+&(\bm{k}_1\leftrightarrow \bm{k}_2,\bm{\hat{\varepsilon}}_{\lambda_1}\leftrightarrow \bm{\hat{\varepsilon}}_{\lambda_2},\omega_1 \leftrightarrow \omega_2 ) \nonumber ,
\end{eqnarray}
where a summation is carried out over the complete energy spectrum of the single-electron intermediate states $\ketm{n}$. Employment of the independent particle approximation allows us to turn from many-electron wavefunctions to one-electron ones. Further simplification of the one-electron transition amplitude can be achieved using the multipole decomposition of the photon field $\bm{A}_{\lambda}(\omega)$ into spherical tensors \cite{Varshalovich/Book:1988}

\begin{eqnarray}
\label{MultipoleExpansion}
\bm{A}_{\lambda}(\omega)=
	4\pi \sum_{J M p} i^{J-p}
	[\bm{\hat{\varepsilon}}_{\lambda}\cdot\bm{Y}_{JM}^{(p)*}(\bm{\hat{k}})]
	\bm{a}^{(p)}_{JM}(\bm{r}),
\end{eqnarray}
where $\bm{Y}_{JM}^{(p)}$ is a vector spherical harmonics and the index $p$ describes the electric ($p=1$) and magnetic ($p=0$) components of the electromagnetic field. In addition, we also perform an expansion of the continuum electron wavefunction into its  partial waves \cite{Eichler/PR:2007}

\newpage
\begin{eqnarray}
\label{PartialWaveExpansion}
\ketm{\mathbf{p}_e m_e}&=&
	\displaystyle{\frac{1}{\sqrt{\varepsilon_e |\bm{p}_e|}}}
	\sum_{jm_j}\sum_{lm_l} i^{l}
	e^{-i \Delta_{jl}} \hspace{3cm}\\ \nonumber
	&\times& \sprm{l,m_l,1/2, m_e}{j,m_j}
	\ketm{\varepsilon_e j l m_j}	
	Y^*_{l m_l}(\hat{p}_e),
\end{eqnarray}
with $\varepsilon_e=\sqrt{\bm{p}_e^2+m^2}$ being the electron energy, $\Delta_{jl}$ the phase factor \cite{Eichler/PR:2007} and $Y^*_{l m_l}(\hat{p}_e)$ the spherical harmonics that depends specifically on the direction of the emitted electron. In the expansion, the summation runs over all total and orbital angular momentum quantum numbers $j$ and $l$, and $\ketm{\varepsilon_e j l m_j}$ are partial waves of the free electron with well-defined electron energy $\varepsilon_e$ and quantum numbers $j,l$, and $m_j$.

The transition amplitude $M^{\lambda_1 \lambda_2}_{J_i M_i J_f M_f m_e}$ from equation (\ref{SimplifiedTransitionAmplitude}) can be further simplified using the expansions (\ref{MultipoleExpansion}) and (\ref{PartialWaveExpansion}). Moreover, by assuming two identical photons, we can write their momenta as $\bm{k}_1=\bm{k}_2=\bm{k}$ and polarization vectors as $\bm{\hat{\varepsilon}}_{\lambda_1}=\bm{\hat{\varepsilon}}_{\lambda_2}=
\bm{\hat{\varepsilon}}_{\lambda}$. Then by choosing $\bm{\hat{k}}$ as the quantization axis, the dot product of the polarization vector and the spherical harmonics in the multipole expansion (\ref{MultipoleExpansion}) can be written as $\bm{\hat{\varepsilon}}_{\lambda}\cdot\bm{Y}_{JM}^{(p)}(\bm{\hat{k}})=\sqrt{[J]/8\pi}(-\lambda)^{p}\delta_{\lambda M}$, where $[J]=2J+1$. Furthermore, by employing the Wigner-Eckart theorem, the amplitude (\ref{SimplifiedTransitionAmplitude}) can be expressed in terms of the reduced transition amplitude, which describes the two-photon interaction with the electron independently of the magnetic quantum numbers $m_a, m_n$, and $m_j$. By carrying out all the above simplifications, we can express the many-electron two-photon amplitude (\ref{GeneralTransitionAmplitude}) within the independent particle approximation by

\begin{widetext}
\begin{eqnarray}
\label{FinalTransitionAmplitude}
M^{\lambda_1 \lambda_2}_{J_i M_i J_f M_f m_e}&=& \nonumber 
	\sum_{p_1 J_1} \sum_{p_2 J_2} \sum_{n_n j_n l_n m_n}
	i^{J_1-p_1+J_2-p_2} 	\sqrt{\frac{[J_1, J_2]}{[j_n, j_a]}} 
	(-\lambda_1)^{p_1} (-\lambda_2)^{p_2}
	\sum_{jm_j} \sum_{lm_l}(-i)^l e^{i\Delta_{jl}} 
	\sprm{j,m_j, l, m_l}{1/2 , m_e}\\ \nonumber 
	&\times& Y_{l,m_l}(\hat{p}_e)  (-1)^{j-m_j} 
	\sprm{j,m_j,J_1,-\lambda_1}{j_n,m_n}
	\sum_{m_a} \sprm{j_a, -m_a, J_i, M_i}{J_f, M_f}
	\\ &\times& \sprm{j_n, m_n, J_2,-\lambda_2}{j_a, m_a}
	\frac{
		\redmem{\varepsilon_e j l}
				{\bm{\alpha} \cdot \bm{ a}^{(p_2)}_{J_2M_2}}
				{n_n j_n l_n}
		\redmem{n_n j_n l_n}
				{\bm{\alpha} \cdot \bm{ a}^{(p_1)}_{J_1M_1}}
				{n_a j_a l_a}
		}{E_{n_a j_a}+\omega_1-E_{n_n j_n}}.
\end{eqnarray}
\end{widetext}

Having the two-photon transition amplitude for the interaction of the atom with the radiation field, we can employ the density matrix theory to obtain the corresponding two-photon ionization cross section. Here, the density matrix of the overall system (ion + outgoing electron) is applied to deal efficiently with the degrees of freedom of the two subsystems and to easily trace-out all those degrees, which are not observed experimentally.
\subsection{Total cross section}
\label{Subsec.TotalCrossSection}

The density matrix of the final system state contains complete information about both the singly ionized atom and the free electron, and can be expressed in terms of the transition amplitude (\ref{FinalTransitionAmplitude}) is given by

\begin{eqnarray}
\label{FinalDensityMatrix}
&&\mem{\alpha_f J_f M_f, \bm{p}_e m_e}{\hat{\rho_f}}{\alpha_f J_f M_f',  \bm{p}_e m_e'}  \nonumber \\ \nonumber
	&&=\sum_{M_i \lambda_1 \lambda_2}
		\sum_{M_i' \lambda_1' \lambda_2'}
		\mem{\alpha_i J_i M_i, \bm{k} \lambda_1 \bm{k} \lambda_2}{\hat{\rho}}{\alpha_i J_i M_i', \bm{k} \lambda_1' \bm{k} \lambda_2'}
		\\
	&&\times 
		M^{\lambda_1 \lambda_2}_{J_i M_i J_f M_f m_e}
		M^{\lambda_1' \lambda_2'*}_{J_i M_i' J_f  M_f' m_e'},
\end{eqnarray}
where $\mem{\alpha_i J_i M_i, \bm{k} \lambda_1 \bm{k} \lambda_2}{\hat{\rho}}{\alpha_i J_i M_i', \bm{k} \lambda_1' \bm{k} \lambda_2'}$ refers to the density matrix of the initial state of the system. As the atom and the incident radiation are initially independent, the initial-state density matrix can be written as a direct product of the neutral atom and the two photon density matrices as follows \cite{Blum/Book:1981}

\begin{eqnarray}
\label{InitialDensityMatrix}
&&\mem{\alpha_i J_i M_i, \bm{k} \lambda_1 \bm{k} \lambda_2}{\hat{\rho}}{\alpha_i J_i M_i', \bm{k} \lambda_1' \bm{k} \lambda_2'}\hspace{1.8cm}\\\nonumber &&=
\mem{\alpha_i J_i M_i}{\hat{\rho_i}}{\alpha_i J_i M_i'}
\mem{\bm{k}\lambda_1}{\hat{\rho}_{\gamma}}{\bm{k}\lambda_1'}
\mem{\bm{k}\lambda_2}{\hat{\rho}_{\gamma}}{\bm{k}\lambda_2'}.
\end{eqnarray}
Here, the $\mem{\bm{k}\lambda}{\hat{\rho}_{\gamma}}{\bm{k}\lambda'}$ are the photon helicity density matrices which allow us to conveniently parametrize the polarization of the photons by means of Stokes parameter 

\begin{eqnarray}
\mem{\bm{k}\lambda}{\hat{\rho}_{\gamma}}{\bm{k}\lambda'}=\frac{1}{2}\twobytwo{1+P_3}{P_1-iP_2}{P_1+iP_2}{1-P_3}.
\end{eqnarray}
In this formalism, it is indeed easy to express any degree of polarization with the linear ($P_1, P_2$) and circular ($P_3$) Stokes parameters and to calculate the corresponding total cross section. As mentioned before, we assume equal momenta of the two photons, however, the photon helicities $\lambda$ (spin projections onto the $\bm{\hat{k}}$ direction) may still differ. Below, we shall assume that the atom is initially unpolarized and that the density matrix of the neutral atom is simply given by

\begin{eqnarray}
\label{InitialDensityMatrix}
\mem{\alpha_i J_i M_i}{\hat{\rho}_i}{\alpha_i J_i M_i'} = \frac{1}{[J_i]}\delta_{M_i M_i'}.
\end{eqnarray}

To extract the observable quantity from the density matrix (\ref{FinalDensityMatrix}), we can define a (so called) "detector operator" $\hat{P}$ which characterizes the experimental detector system as a whole. This operator determines the probability for an "event" to be recorded at the detector. Then, the probability is simply given by the trace of the product of the detector operator and the density matrix. 
Here, we consider an electron detector insensitive to the electron polarization detecting electrons in $4\pi$ solid angle. The detector can be thus described by the operator $\hat{P}=\int d\hat{p}_e \sum_{m_e} \ketm{\bm{p}_e m_e}\bram{\bm{p}_e m_e}$. Moreover, as we do not observe the final ionic state, we have to sum over the corresponding quantum numbers $J_f$ and $M_f$. Then, the total cross section for non-resonant ionization of an atom by two photons with $\bm{k}_1=\bm{k}_2=\bm{k}$ and $\bm{\hat{k}}||\bm{\hat{z}}$ is given by

\begin{eqnarray}
\label{TotalCrossSection}
\sigma(\omega)&=&\frac{32 \pi^5 \alpha^2}{\omega^2} \sum_{J_f M_f}\mathrm{Tr}(\hat{P} \hat{\rho}_f)\\\nonumber
	&=&\frac{32 \pi^5 \alpha^2}{\omega^2} \frac{1}{[J_i]}
	\sum_{ \lambda_1 \lambda_2 \lambda_1' \lambda_2'}
	\mem{\bm{k}\lambda_1}
		{\hat{\rho}_{\gamma}}
		{\bm{k}\lambda_1'}
	\mem{\bm{k}\lambda_2}
		{\hat{\rho}_{\gamma}}
		{\bm{k}\lambda_2'}\\\nonumber &\times&
	\int d\hat{p}_e \sum_{J_f M_i M_f m_e} 
	M^{\lambda_1 \lambda_2}_{J_i M_i J_f M_f m_e}
	M^{\lambda_1' \lambda_2'*}_{J_i M_i J_f M_f m_e}.
\end{eqnarray}
As this expression represents second-order cross section, it has the units of $[L^4T]$.

\section{Computations}
\label{Sec.Computation}

From the theoretical description above, it can be seen that the main computation challenge lies in the infinite summations of the reduced matrix elements (\ref{FinalTransitionAmplitude}) over all multipole orders and infinite number of intermediate states. To deal with this numerically, the infinite summations over the multipoles of each of the two photons were restricted to a maximum value of $J_{\mathrm{max}}=5$. This limit is sufficient to obtain convergence of the corresponding total cross section at less than 0.001\% level. To sum over the infinite number of intermediate states, finite basis-set \cite{Sapirstein/JPB:1996} constructed from $B$-splines by applying  the dual-kinetic-balance approach \cite{Shabaev/PRL:2004} was employed. This technique allows us to reduce infinite sum over the intermediate states in (\ref{FinalTransitionAmplitude}) to finite sum over pseudospectrum. This approach has been previously successfully applied, for example, in the calculations of two-photon decay rates of heliumlike ions \cite{Volotka/PRA:2011, Surzhykov/PRA:2010} or cross sections of x-ray Rayleigh scattering \cite{Volotka/PRA:2016}. The continuum-state wavefunctions were obtained by numerical solutions of the Dirac equation with help of the RADIAL package \cite{Salvat/CPC:1995}.

In order to account for the screening effects, we solve the Dirac equation with a screening potential, which partially accounts for the interelectronic interaction. We use the core-Hartree potential, which corresponds to a potential created by all bound electrons except of the active electron. The core-Hartree potential reproduces the electron binding energies in excellent agreement to the experimental values within $\pm 0.2\%$ error for all atoms under consideration.
To analyse the sensitivity to the choice of potential, in addition to the core-Hartre potential, two different screening potentials were also used. The potential taken from Ref. \cite{Salvat/PRA:1987}, to which we refer as the "Salvat" potential and Salvat potential modified in a way to reproduce experimental binding energies $E_{\mathrm{bind}}^{\mathrm{exp}}$, referred to as "Salvat $E_{\mathrm{bind}}^{\mathrm{exp}}$". All results presented were calculated using the core-Hartree potential, except for Fig. \ref{Fig.PotentialComparison} where results from the different potentials are compared. 

In addition, in order to check the consistency of our results, we carried out the calculations in length and velocity gauges. The results for both gauges were in a perfect agreement as expected for any local potential. Even though the agreement of the two gauges does not prove validity of the results, it shows that the effective single-electron amplitudes (\ref{FinalTransitionAmplitude}) are properly implemented in our codes.
%
%
%
\section{Results and discussion}
\label{Sec.ResultsAndDiscussion}

\begin{figure*}
\centering
\begin{minipage}{.5\textwidth}
  \centering
  \includegraphics[width=0.97\linewidth]{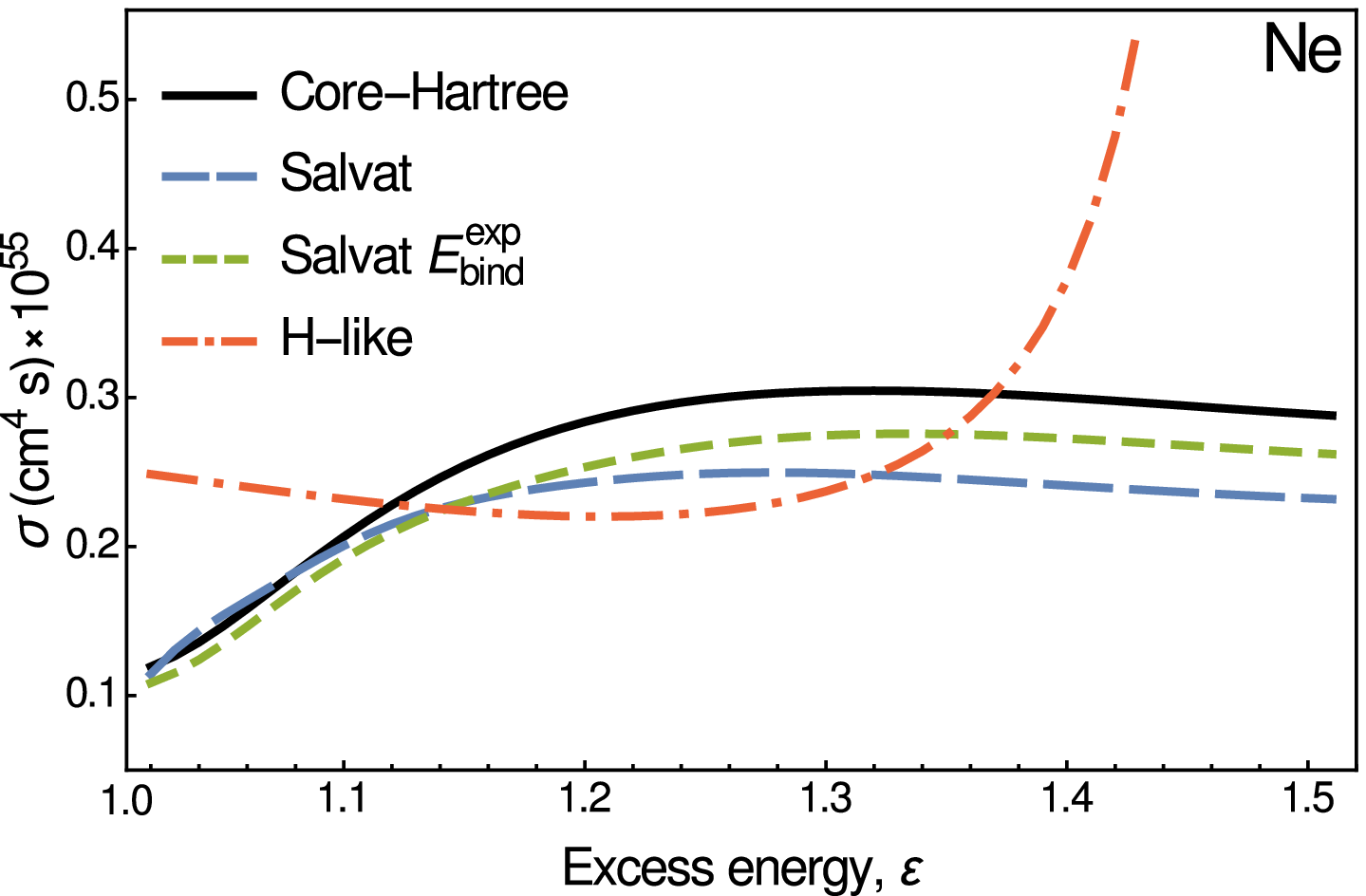}
  \end{minipage}%
\begin{minipage}{.5\textwidth}
  \centering
  \includegraphics[width=0.97\linewidth]{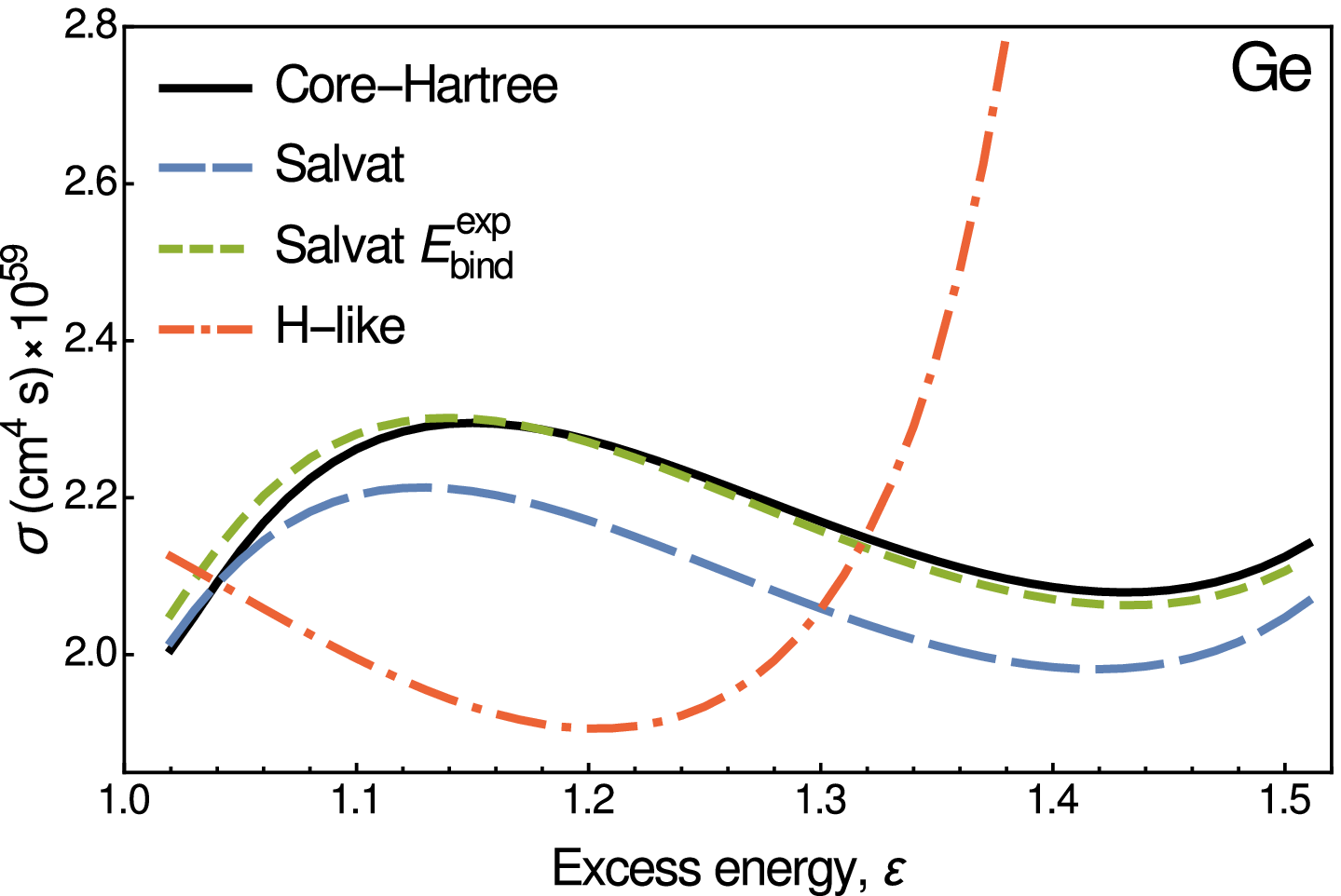}
   \end{minipage}
\begin{minipage}{.5\textwidth}
\centering
  \includegraphics[width=0.97\linewidth]{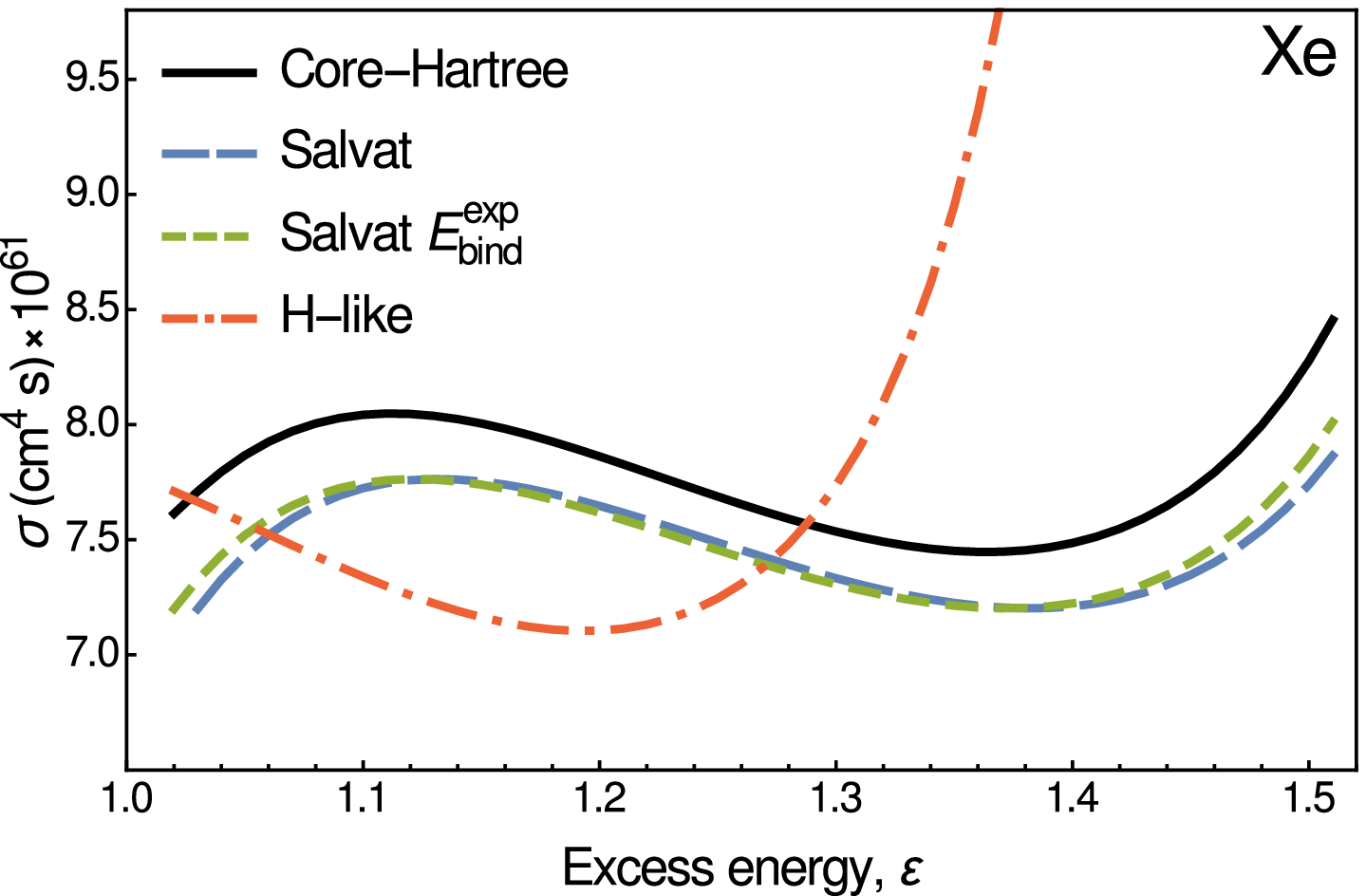}
  \end{minipage}%
\begin{minipage}{.5\textwidth}
  \centering
  \includegraphics[width=0.97\linewidth]{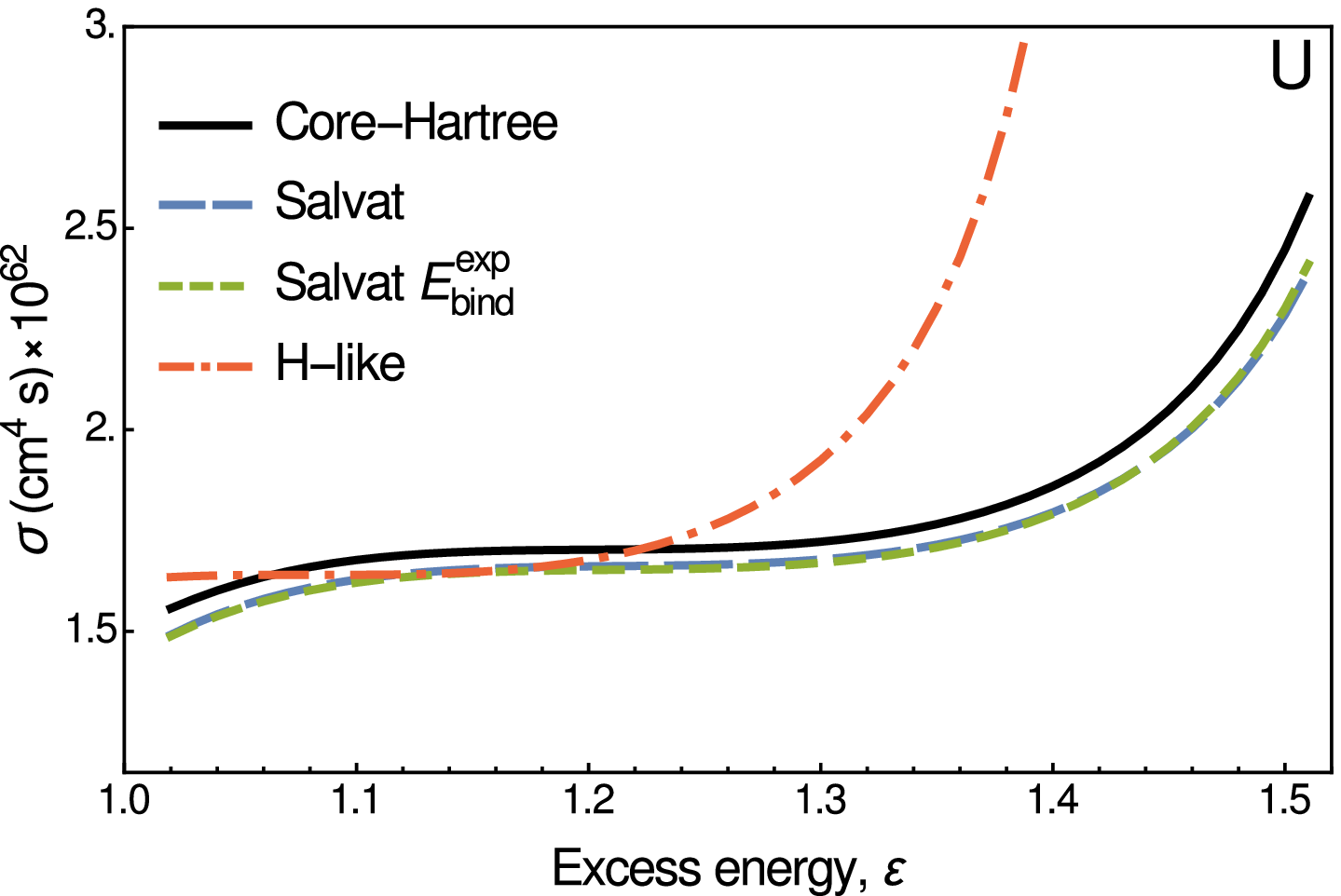}
   \end{minipage}
\caption{The total two-photon $K$-shell ionization cross section $\sigma$ as a function of excess energy $\varepsilon$ (in units of $K$-shell binding energy) for ionization by linearly polarized photons. Results are shown for the two-photon ionization of hydrogenlike (dash-dotted red) ions and neutral atoms calculated in three different potentials: core-Hartree (solid black), Salvat (long-dashed blue) as well as Salvat $E_{\mathrm{bind}}^{\mathrm{exp}}$ (short-dashed green) potentials. Calculations were performed for neon, germanium, xenon, and uranium atoms (as labelled).}\label{Fig.PotentialComparison}

\end{figure*}

Even though the formalism derived in Sec. \ref{Sec.Theory} applies generally for neutral atoms as well as ions, detailed calculations have been carried out for $K$-shell two-photon ionization of neutral neon, germanium, xenon, and uranium atoms. Specifically, the contributions of relativistic and screening effects to the total cross section have been investigated. In this section, we will compare cross sections for $K$-shell ionization of hydrogenlike and neutral atoms. Since there are two electrons in the $K$-shell of neutral atoms but only one electron in the $K$-shell hydrogenlike ions, we introduce an additional factor of two in the hydrogenlike calculation for the sake of comparison.

We begin by comparing the ionization of hydrogenlike and neutral atoms in terms of the total cross section as a function of so called excess energy. Excess energy represents the factor by which the combined photon energy exceeds the ionization threshold, i.e., $\varepsilon=2\omega/E_{\mathrm{bind}}$. Figure \ref{Fig.PotentialComparison} presents the total cross sections for ionization of neutral (solid black) as well as for H-like (dashed-dotted red) neon, germanium, xenon, and uranium by linearly polarized photons. We can notice that the first resonant behaviour in the total cross section occurs in lower energy for H-like ions than for neutral atoms. This resonant behaviour occurs when the single photon energy reaches the $1s \rightarrow 2p$ transition energy. Although the $2p$ state is generally occupied for neutral atoms, the resonant two-photon ionization can be understood as follows. The $2p$ electron is ionized by the first photon and the corresponding vacancy is then filled by excitation of the $1s$ electron by the second photon. Since the present work is devoted to the non-resonant ionization, the $1s \rightarrow 2p$ resonant energy and the ionization threshold define the energy range of current interest. 

The more significant difference between neutral and H-like systems lies in the decrease of the total cross section near the ionization threshold. This cross section reduction is strongest for elements with nuclear charge $Z=7-12$ and becomes much less significant for heavy atoms. In the case of H-like ions, no such behaviour has been predicted and the total two-photon ionization cross section is slowly decreasing in non-resonant energy regions \cite{Koval/Dissertation, Koval/JPB:2003}, which we also confirm by the present calculations. This means that the change of the total cross section for light neutral elements close to the ionization threshold can be directly linked to the deviation of the binding potential from the Coulomb potential created by the nucleus. In the next subsection, we will investigate these effects (which we refer to as "screening effects") further by looking at the $s-$ and $d-$ partial waves of the free electron. These partial waves strongly dominate others as they are the only allowed by two electric dipole ($E1E1$) transitions.

\begin{figure*}
\centering
\begin{minipage}{.5\textwidth}
  \centering
  \includegraphics[width=0.97\linewidth]{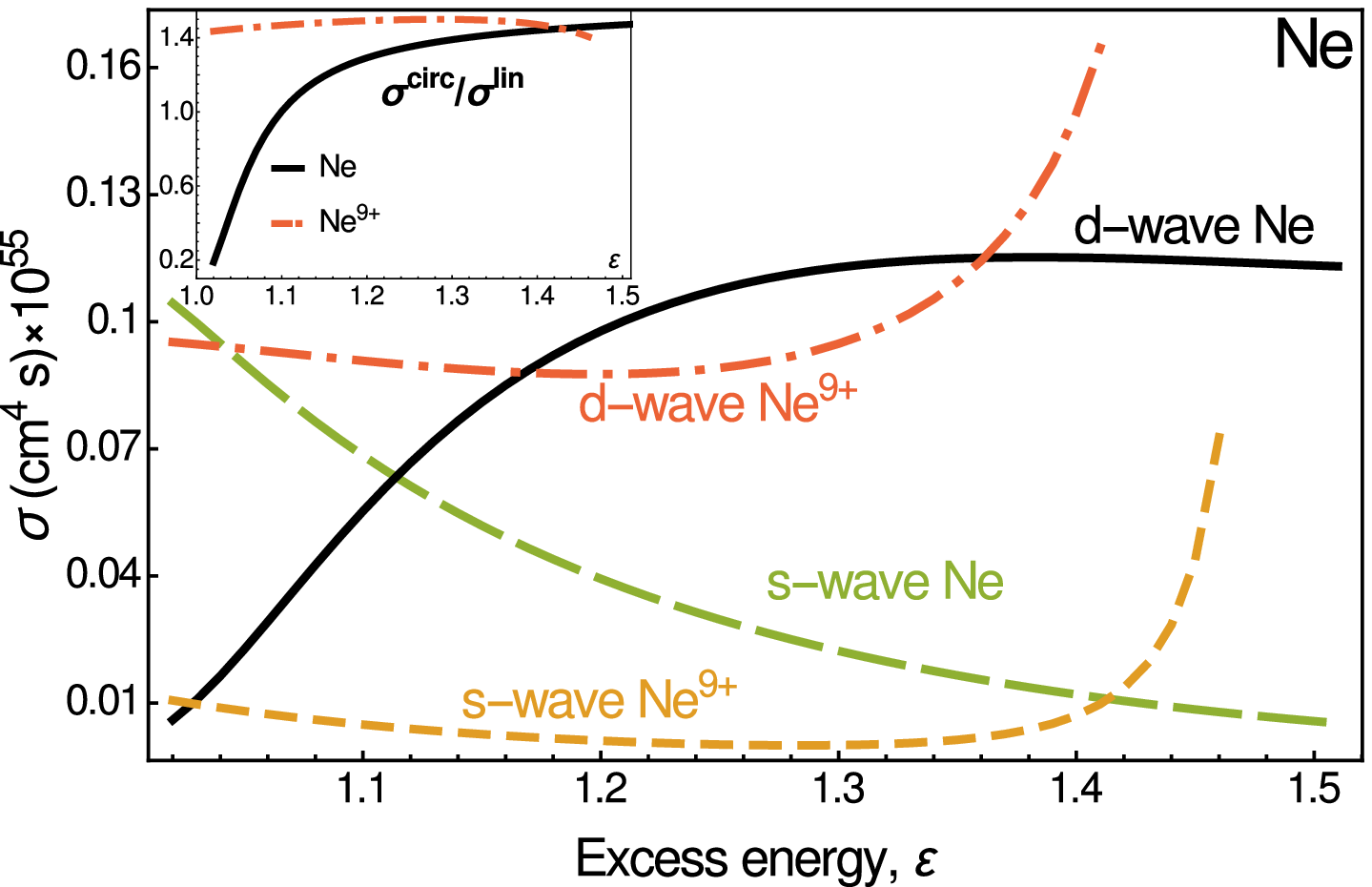}
  \end{minipage}%
\begin{minipage}{.5\textwidth}
  \centering
  \includegraphics[width=0.97\linewidth]{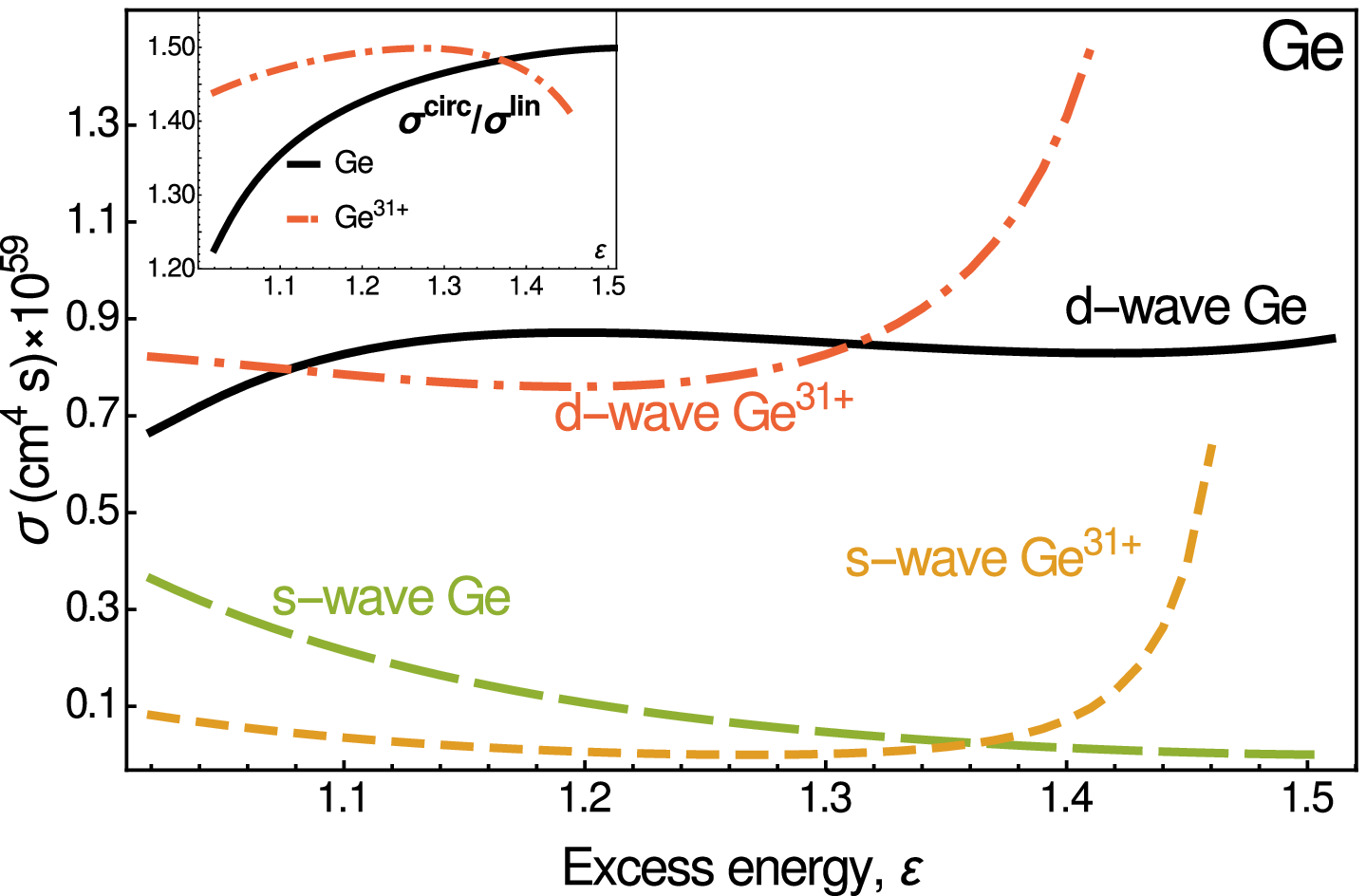}
   \end{minipage}
\begin{minipage}{.5\textwidth}
\centering
  \includegraphics[width=0.97\linewidth]{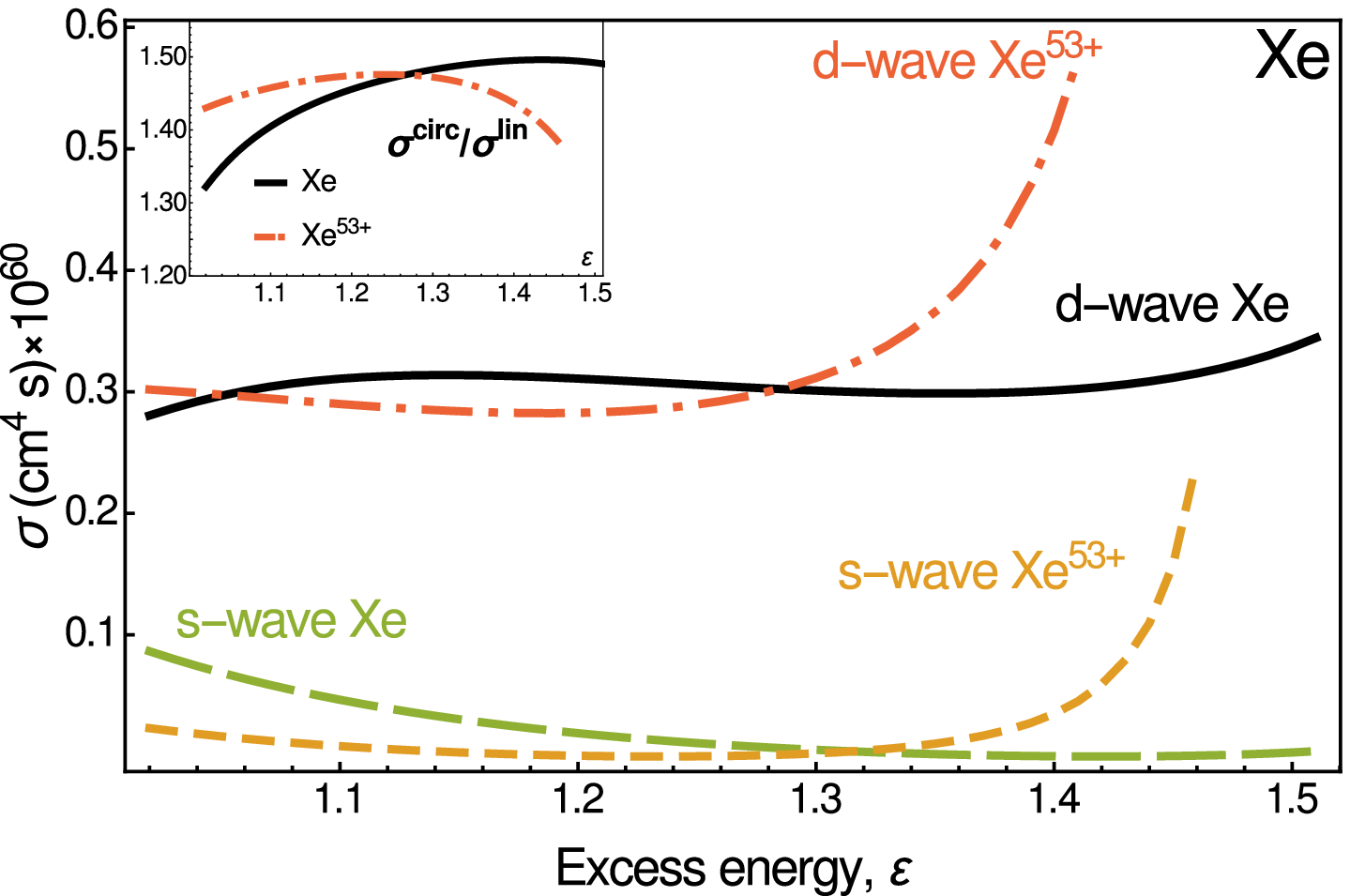}
  \end{minipage}%
\begin{minipage}{.5\textwidth}
  \centering
  \includegraphics[width=0.97\linewidth]{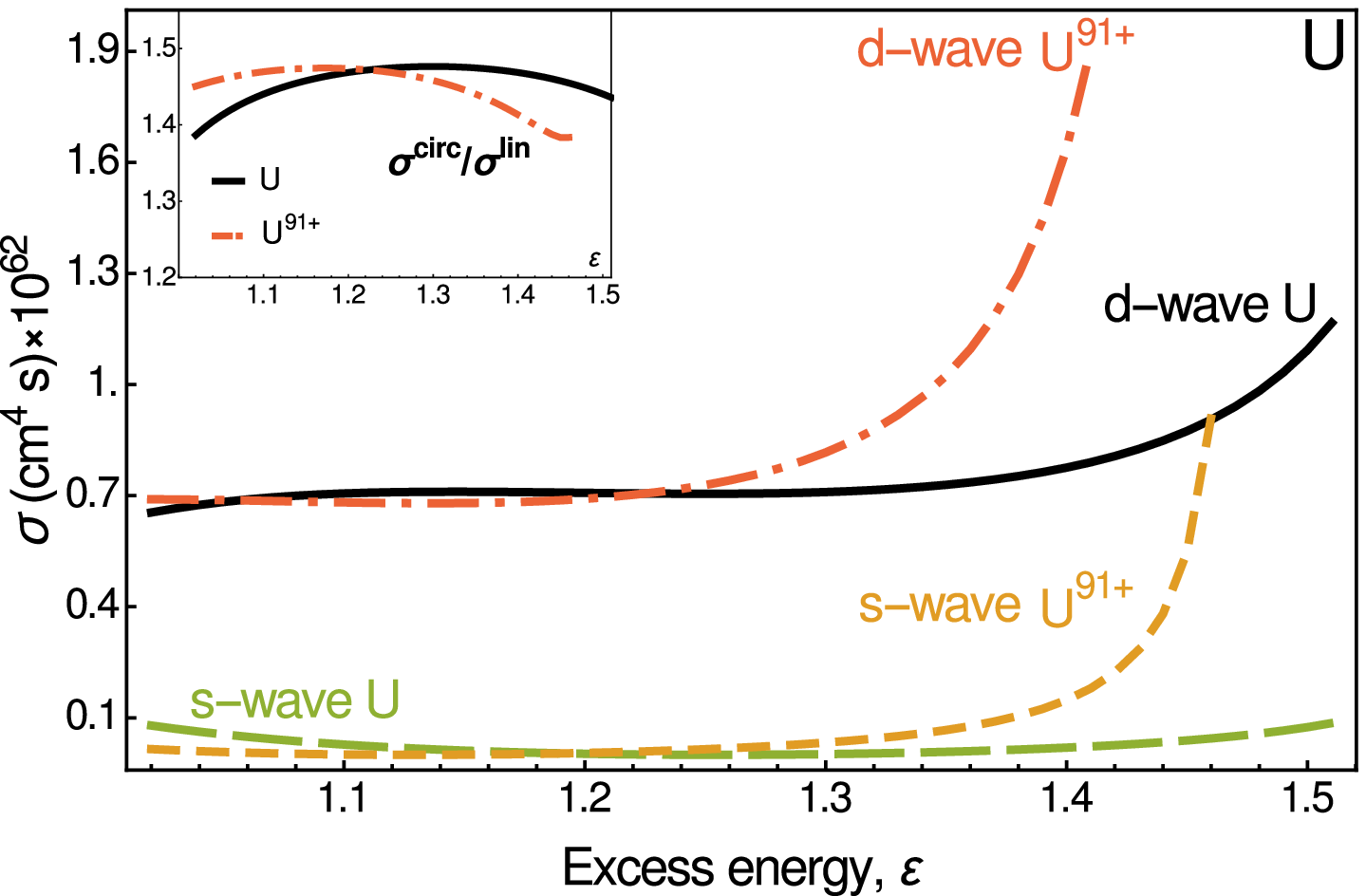}
   \end{minipage}
\caption{The partial-wave cross section as a function of excess energy compared for the $s\rightarrow p\rightarrow s$ (long-dashed green) and $s\rightarrow p\rightarrow d$ (solid black) ionization channels of neutral atoms by linearly polarized light. Results for the $s\rightarrow p\rightarrow s$ (short-dashed orange) and $s\rightarrow p\rightarrow d$ (dash-dotted red) ionization channels of H-like ions are also shown. As a consequence of screening effects, the $s\rightarrow p\rightarrow s$ channel becomes dominant for low-$Z$ atoms in near-threshold photon energies. The ratio of cross sections corresponding to ionization by circularly and linearly polarized light is presented in top left corner of each figure. The screening effects also result in a deviation of this ratio from the known estimate $ \sigma^{\mathrm{circ}}/  \sigma^{\mathrm{lin}} \approx 3/2$. }\label{Fig.PartialWave}
\end{figure*}

Figure \ref{Fig.PotentialComparison} shows also the comparison between the three screening potentials (solid black, dashed green, and blue curves) introduced in Sec. \ref{Sec.Computation}. We see that for low-$Z$ and medium-$Z$ atoms, the core-Hartree and Salvat potentials differ in the magnitudes of the total cross section by less than $25\%$. This is partially caused by the calculated value of the binding energy. When the Salvat potential was modified to reach perfect agreement with the experimental binding energies, the cross section difference from the core-Hartree calculation was reduced to about $10\%$. Therefore, even though part of the difference between the cross sections as predicted by each potential arises from the difference of binding energies, the distinct potential formulations also result in a deviation. Despite the small magnitude differences, all screening potentials predict similar energy dependence of the total cross section. The agreement of these potentials justifies that the obtained behaviour and magnitude is not very sensitive to the choice of potential. We ascribe the difference between the core-Hartree and Salvat calculations as an uncertainty of presented results. The uncertainty decreases from 25$\%$ for Ne to 10$\%$ for U. We restrict all further discussion to the use of core-Hartree potential.

\subsection{Partial wave analysis}\label{Subsec.PartialWave}

To gain deeper understanding of the total cross section results, we now wish to look at the dominant $E1E1$ ionization channels. In this approximation, only the $J=1$ multipole of each of the two photons is considered. Since we are interested in ionization of a $1s$ electron with zero orbital angular momentum $l=0$, $E1E1$ transition allows only two possible ionization channels; $s\rightarrow p \rightarrow s$ and $s \rightarrow p \rightarrow d$. Therefore, only $s-$ and $d-$ partial waves of the free electron are allowed in dipole approximation. While both of these channels are open for linear and unpolarized light, only the $s\rightarrow p \rightarrow d$ channel is open for circularly polarized light. This restriction comes from the conservation of the angular momentum projection. Since we are considering two equally circularly polarized photons, the angular momentum projection must change by $\pm 2$, then $|m_j|>1/2$ is always the case, making the final $s-$state forbidden. This is a point worth remembering. As we will soon see, the absence of the $s\rightarrow p\rightarrow s$ channel leads to a magnification of the screening effects, which increases the probability of experimental detection of these effects.

Figure \ref{Fig.PartialWave} shows the plots of the partial-wave cross sections considering only the $s-$ or $d-$ partial-waves of the continuum electron as a function of excess energy. Results are presented for ionization of H-like (dash-dotted red and short-dashed orange) and neutral (solid black and long-dashed green) neon, germanium, xenon, and uranium atoms by linearly polarized light. We can see that the energy dependence of the partial-wave cross section of H-like ions fulfils our expectation we gained from Fig. \ref{Fig.PotentialComparison}. The cross sections of $s\rightarrow p\rightarrow d$ channel always dominates the $s\rightarrow p\rightarrow s$ channel and the two curves remain approximately parallel. Analogously to the total cross section, both channels can be considered constant in non-resonant energy region up to the proximity of the $1s\rightarrow 2p$ resonant energy. Similar behaviour can be seen in the case of neutral uranium. However, for neutral atoms with lower nuclear charge, we observe a competition of the two partial waves in near-threshold energy region. A drop of the dominant channel occurs and creates a minimum of the cross section, analogous to the Cooper minimum in single photon ionization process. The minimum is most pronounced for neon, for which the cross section of the $s\rightarrow p\rightarrow s$ channel is greater than the dominant $s\rightarrow p\rightarrow d$ channel in an energy region from the ionization threshold up to a crossing point of the channels at $\varepsilon=1.12$. This crossing of the ionization channels is present for atoms with nuclear charges in the range $Z=5-13$. Although for elements in this range other than neon, the crossing point lies in lower energies and the effects are thus weaker.

In the top left part of each of the figures \ref{Fig.PartialWave}, the ratio of total cross section for ionization by circularly $\sigma^{\mathrm{circ}}$ and linearly $\sigma^{\mathrm{lin}}$ light are also presented. According to the known estimate $\sigma^{\mathrm{circ}} / \sigma^{\mathrm{lin}} \approx 3/2$ \cite{Lambropoulos/PRL:1972}, the ratio should be always approximately equal to $3/2$ in non-resonant energy region. While this holds true for the H-like ions (dash-dotted red curve), in the case of neutral atoms (solid black curve), the screening effects result in a strong deviation from the estimated value. This follows directly from the discussion of partial waves above.  

\subsection{Screening and relativistic effects}\label{Subsec.Effects}

\begin{figure}
\centering
\begin{minipage}{.47\textwidth}
      \includegraphics[width=\linewidth, scale=0.2]{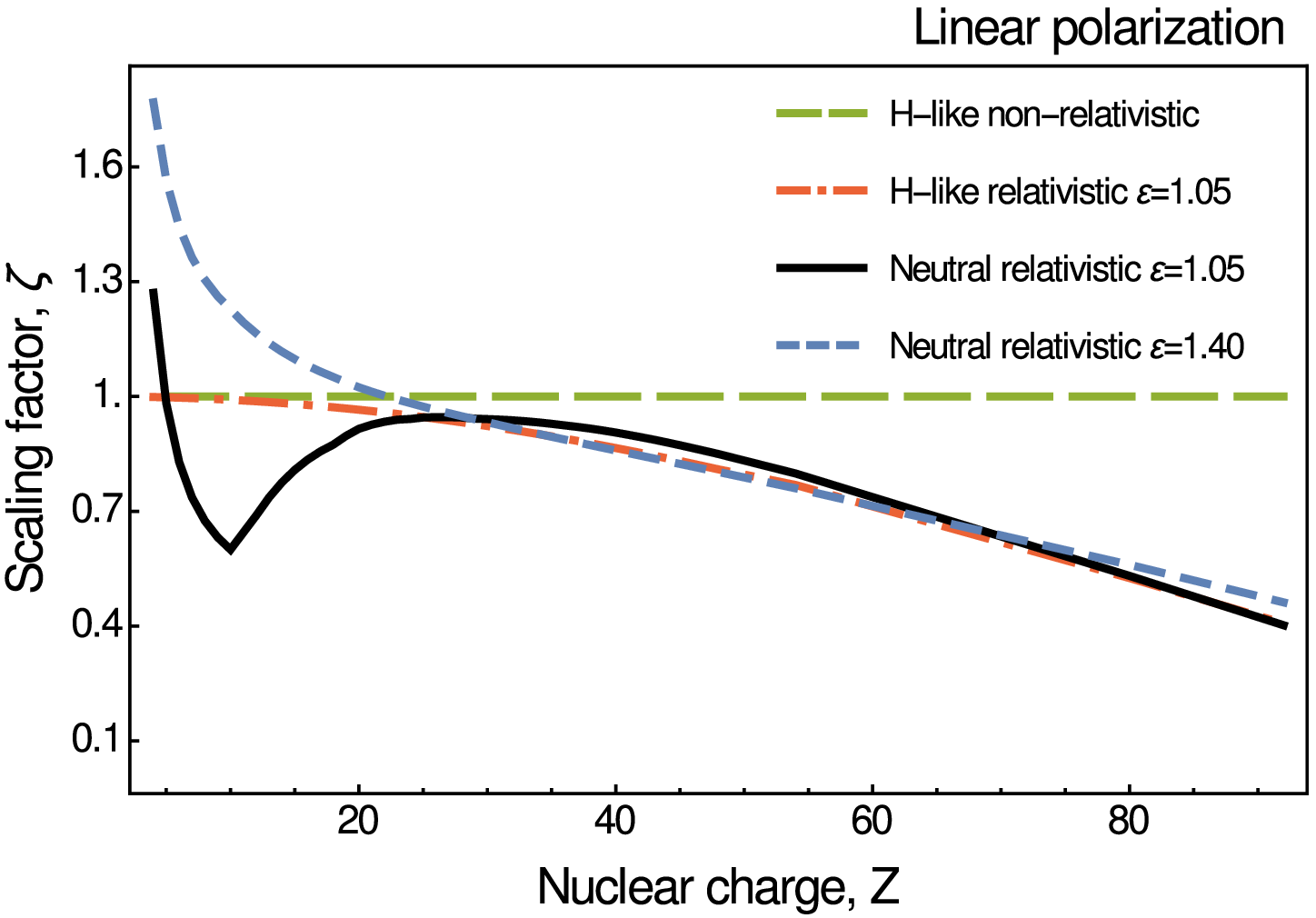} 
     \end{minipage}
\begin{minipage}{.47\textwidth}
      \includegraphics[width=\linewidth, scale=0.2]{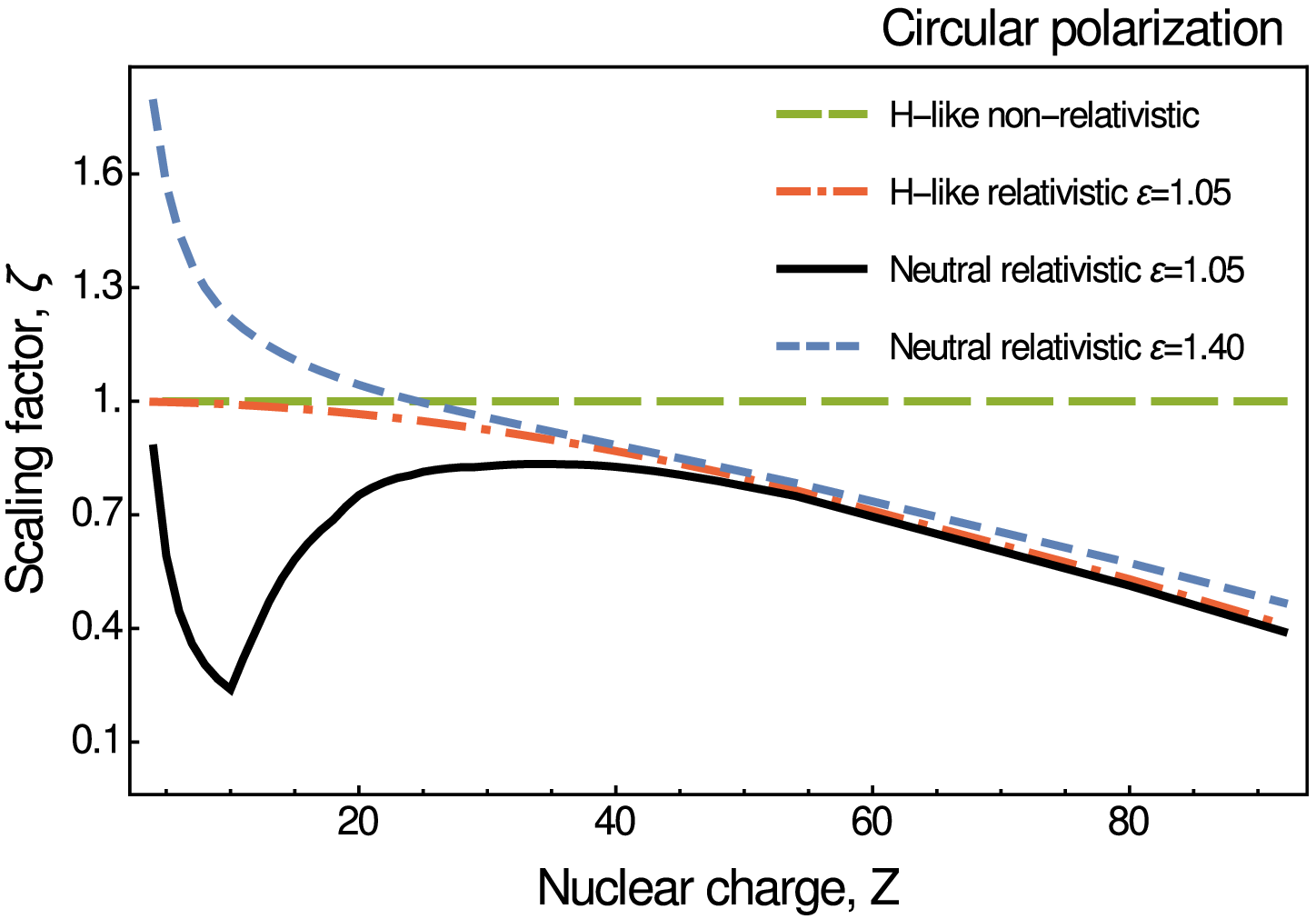} 
     \end{minipage}
\begin{minipage}{.47\textwidth}
      \includegraphics[width=\linewidth, scale=0.2]{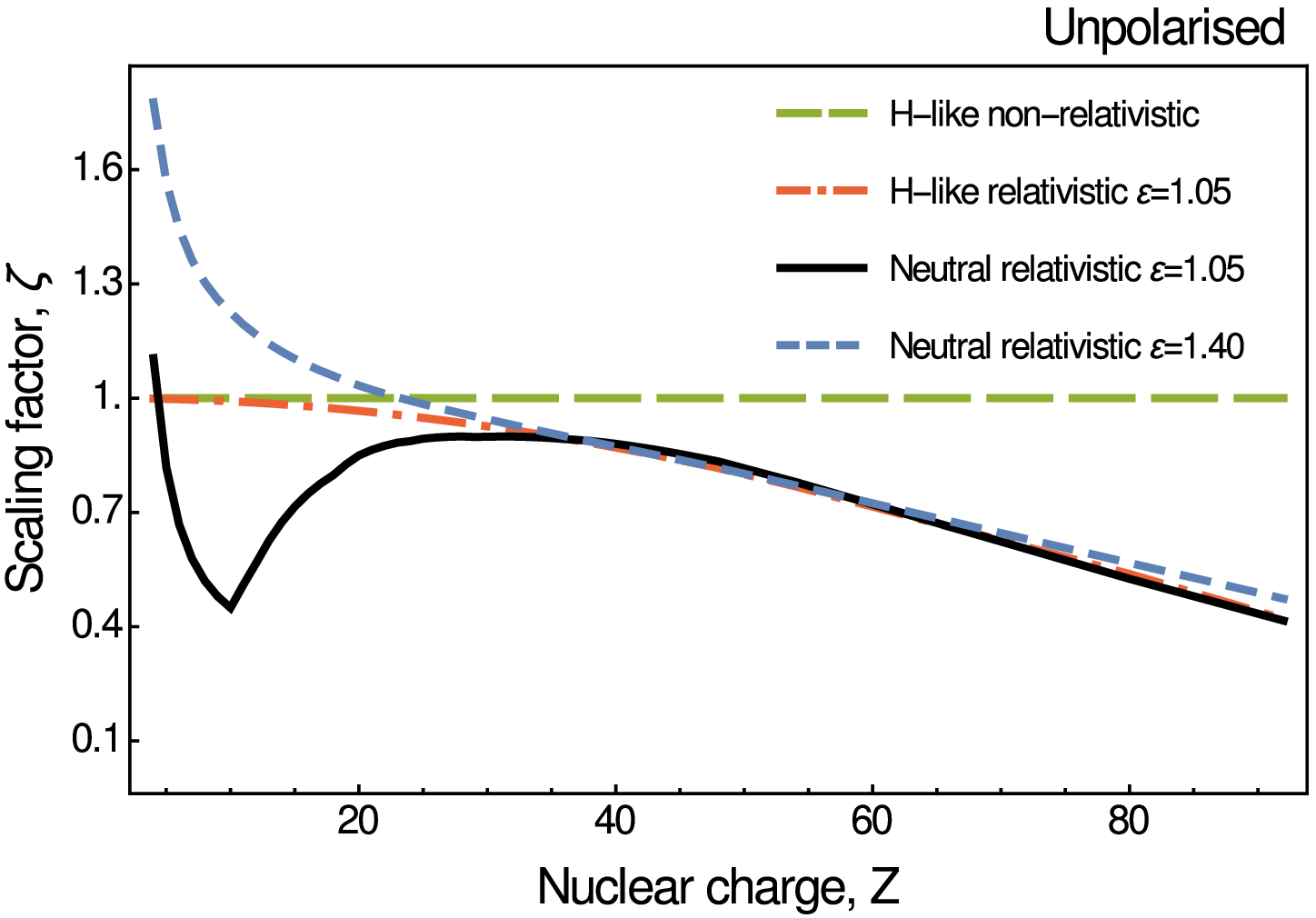} 
    \end{minipage}

\caption{The scaling factor $\zeta$ as a function of nuclear charge for ionization of a $1s$ electron of neutral atoms ($Z=4-92$) by two linearly (top), circularly (middle), and unpolarized (bottom) photons at the excess energies $\varepsilon=1.05$ and $\varepsilon=1.40$. According to the non-relativistic scaling of H-like ions (dashed green), the cross section scales with $Z^{-6}$. The deviation from this scaling due to relativistic effects  is clearly visible for H-like (dash-dotted red) as well as neutral (solid black and short-dashed blue) atoms. Moreover, further deviation of the scaling factor in low-$Z$ region is present for neutral atoms due to screening effects.}\label{Fig.Effects}
\end{figure}

In previous subsections, we already saw that one needs to take screening effects into account for low- and medium-$Z$ elements. Moreover, in Ref. \cite{Koval/JPB:2003} it is shown that in two-photon ionization of H-like ions, the relativistic effects cannot be neglected for heavy atoms. It is, therefore, reasonable to expect similar behaviour for ionization of neutral atoms. It is the purpose of this subsection to show the relative strengths and nuclear charge dependences of both these effects as well as their contributions to the total cross section.

In non-relativistic theory, the non-resonant cross section for the two-photon ionization of H-like ions in dipole approximation scales with the nuclear charge as $\sigma(Z,\omega Z^2)=\sigma(Z=1,\omega)Z^{-6}$ \cite{Zernik/PR:1964}. We will use the same way as in Ref. \cite{Koval/JPB:2003} and introduce so called scaling factor $\zeta$ to the above expression, i.e., $\sigma(Z,\omega Z^2)=\zeta(Z)\sigma(Z=1,\omega)Z^{-6}$. The deviation of the scaling factor from the value $1$ then represents various effects arising from the full relativistic description and/or the interelectronic interaction. For non-relativistic $E1E1$ calculation in Coulomb potential, the scaling factor is $\zeta(Z)=1$ for all $Z$ values and is almost independent of the excess energy in the non-resonant region.

Figure \ref{Fig.Effects} shows the plot of the scaling factor $\zeta (Z)$ as a function of nuclear charge for two-photon ionization by linearly, circularly, and unpolarized light. The results are shown for non-relativistic (dashed green) and relativistic (dash-dotted red) calculations for ionization of H-like ions as well as relativistic calculation for ionization of neutral atoms at $\varepsilon=1.05$ (solid black) and $\varepsilon=1.40$ (long-dashed blue) excess energies. We can see that for neutral atoms, there are two distinct deviations of the scaling factor from the constant non-relativistic value. One of the deviations stretches between the medium- and high-$Z$ region and is also present for the case of hydrogenlike atoms. The second deviation lies in the low-$Z$ region and is present only for the ionization of neutral atoms. Let us start with the deviation in the low-$Z$ region. This deviation results from the interelectronic interaction, which decrease the electron binding energies and as a result, increase the total cross section. We can see, that this is indeed the case for the $\varepsilon=1.40$ excess energy, where the screening effects increase the total cross section in the low-$Z$ region. This increase rapidly weakens with increasing nuclear charge as we would expect. However, for $\varepsilon=1.05$, the screening effects result in decrease of the cross section, with a maximum at $Z=10$. This trough in the scaling factor directly reflects the decrease of cross section we have seen in Figs. \ref{Fig.PotentialComparison} and \ref{Fig.PartialWave}. The sharpness of the trough is a result of the discrete values of the nuclear charge $Z$ values. For photon energies exceeding the ionization threshold by more than $15\%$, i.e. $\varepsilon > 1.15$, the trough disappears. From figure \ref{Fig.Effects}, we can see that the screening effects are strongest for the case of ionization by circularly polarized light. We can understand this from the partial-wave analysis in Sec. \ref{Subsec.PartialWave}. If we look at the partial wave cross sections for neon in Fig. \ref{Fig.PartialWave}, we can see that the dominant $s\rightarrow p \rightarrow d$ channel drops strongly near the ionization threshold. For ionization by linearly and unpolarized light, it is partially balanced by the increase of the $s\rightarrow p\rightarrow s$ channel. However, as explained before, for ionization by circularly polarized light the $E1E1$ transition allows only the dominant $s\rightarrow p \rightarrow d$ channel to be open. Therefore, due to the lack of the  final $s-$ partial wave, the drop of the cross section does not get balanced out and the screening effects become stronger. 

\begin{figure}[h]
\centering
\begin{minipage}{.47\textwidth}
      \includegraphics[width=\linewidth, scale=0.2]{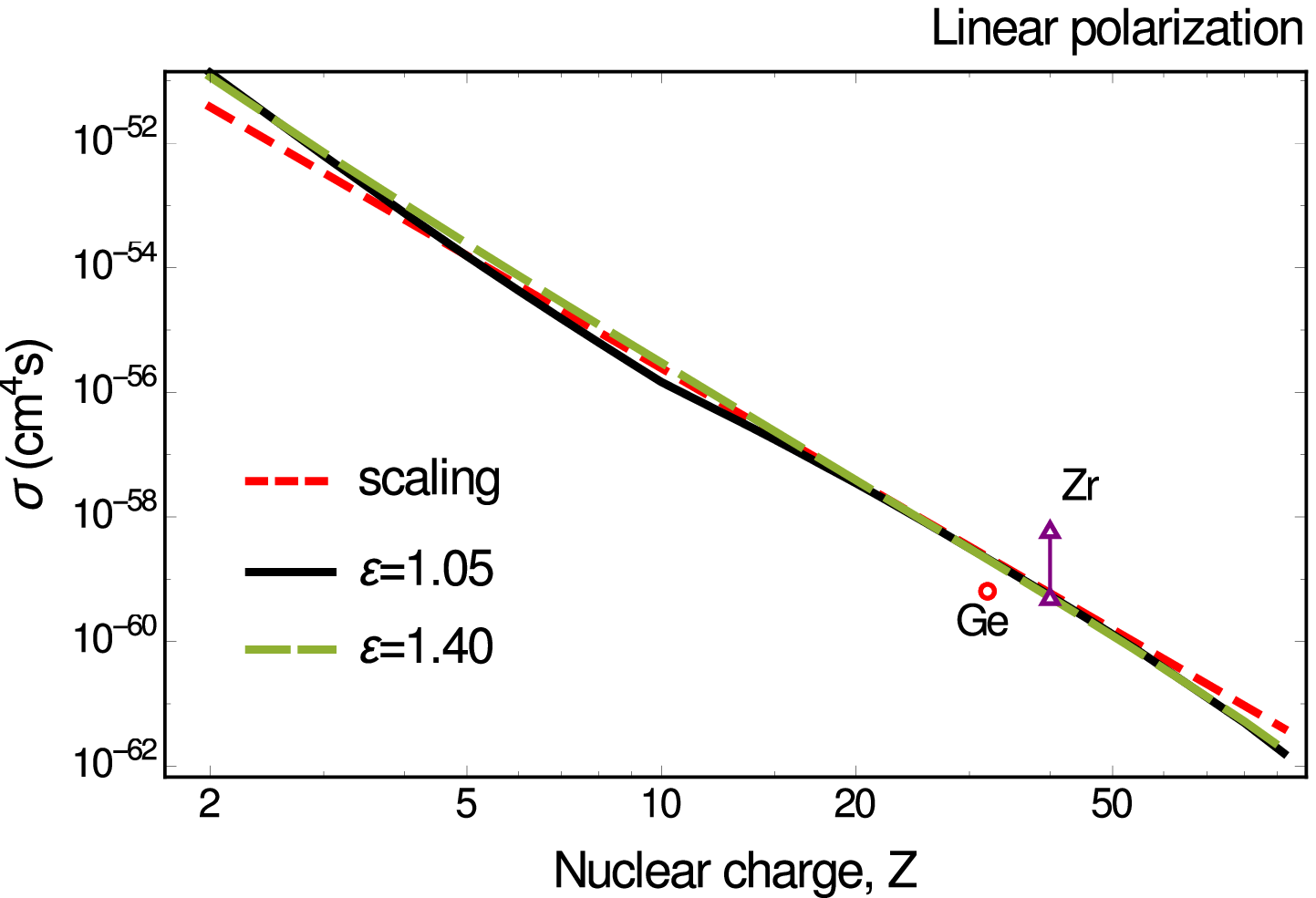} 
     \end{minipage}
\begin{minipage}{.47\textwidth}
      \includegraphics[width=\linewidth, scale=0.2]{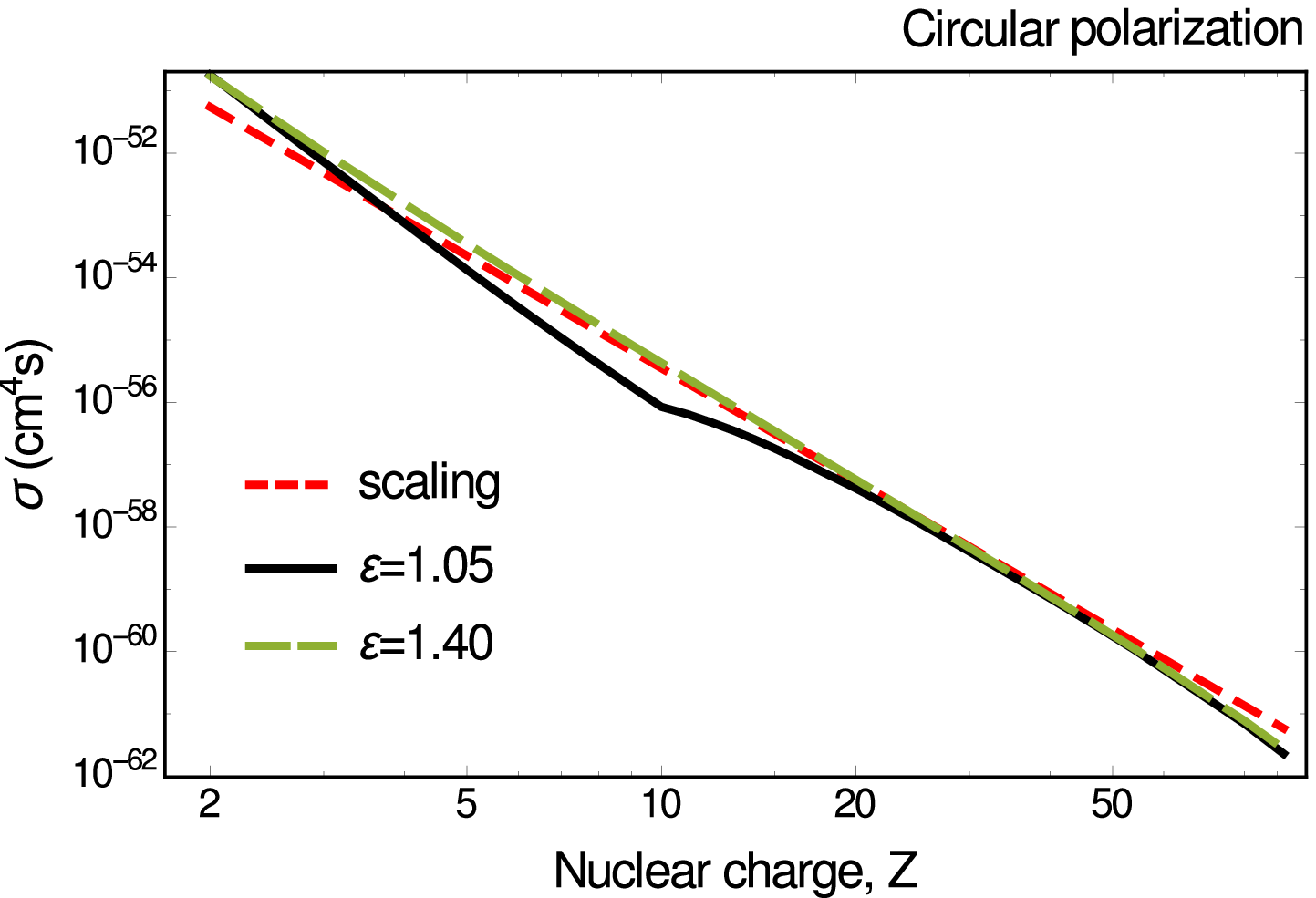} 
     \end{minipage}

\caption{The total cross section for two-photon ionization for linearly (top) and circularly (bottom) polarized light as a function of nuclear charge. The cross section is plotted for two excess energy values, $\varepsilon=1.05$ (solid black) and $\varepsilon=1.4$ (long-dashed green). The $Z^{-6}$ scaling law (short-dashed red) and experimental values for germanium \cite{Tamasaku/NP:2014} and zirconium \cite{Ghimire} atoms are also shown.}\label{Fig.ZScaling}
\end{figure}

The second deviation of the scaling factor in medium- and high-$Z$ region in the Fig. \ref{Fig.Effects} comes from the relativistic effects. The importance of these effects continuously grows with increasing nuclear charge $Z$. We can also see that unlike screening effects, relativistic effects are independent of polarization. This means that relativistic effects influence all partial waves in a same way.  For ionization of uranium by light of any polarization, the relativistic effects decrease the total cross section by about a factor of two. 
We would expect, that the relativistic effects would be stronger for the ionization of hydrogenlike ions than for ionization of neutral atoms, since the electron binding energies of hydrogenlike ions are higher. However, from Fig. \ref{Fig.Effects}, we can see, that the deviation of the scaling factor (and therefore the cross section itself) for neutral atoms due to relativistic effects is similar as for H-like ions.

\subsection{Comparison with experiment}\label{Subsec.Zdependence}

Due to the relativistic and screening effects, corrections to the non-resonant two-photon ionization scaling law $\sigma(Z)=\sigma(Z=1)Z^{-6}$ increase in complexity. The magnitude of these effects depends mainly on the nuclear charge but screening effects also depend on the incident photon energies and polarizations. That is why we present the $Z$-dependence of the total cross section in addition to the scaling factor given in previous subchapter. Figure \ref{Fig.ZScaling} shows calculated cross sections for elements in the range $Z=4-92$ for two energies; $\varepsilon=1.05$ (solid black) and $\varepsilon=1.40$ (long-dashed green) as well as the scaling law (short-dashed red). Total cross sections for other photon energies $1.05 < \varepsilon < 1.40$ lay in between the two corresponding lines in Fig. \ref{Fig.ZScaling}. The cross section difference between the two energies arises due to the screening effects as explained before. Figure \ref{Fig.ZScaling} also shows experimental values for the $K$-shell ionization of neutral Ge and Zr atoms. We can see that our result for Ge is close to the experimental value as well as for Zr, which lies within the experimental uncertainty. However, in another experiment, Doumy, \textit{et al.} \cite{Doumy/PRL:2011} measured the two-photon ionization of heliumlike Ne to be $7\times 10^{-54}$ cm$^4$ s. Theoretical calculations \cite{Sytcheva/PRA:2012, Novikov/JPB:2001, Koval/Dissertation} of this cross section resulted in a discrepant value, lower by about three orders of magnitude. We applied our formalism for the case of Ne$^{8+}$ as well, and obtained a cross section of 3.1$\times 10^{-57}$ cm$^4$ s which is in an agreement with previous calculations \cite{Sytcheva/PRA:2012, Koval/Dissertation, Novikov/JPB:2001}. Thus, the three orders of magnitude deviation obtained suggests a resonant enhancement of the cross section and can be explained by broader spectral bandwidth of the FEL employed. 
\vspace{0.3cm}

%
%
%
%
%
\section{Summary and outlook}
\label{Sec.SummaryAndOutlook}

The non-resonant two-photon ionization of neutral atoms has been described in fully relativistic theory based on second-order perturbation theory and Dirac equation. Using the independent particle approximation and particle-hole formalism, the many-electron transition amplitude describing the electron-photon interaction has been simplified to one-electron amplitude. An expression of the total two-photon ionization cross section has been obtained using the framework of density matrix theory and the transition amplitude. Detailed calculations of the total cross section have been carried out for ionization of neon, germanium, xenon, and uranium atoms using three screening potentials. Our results show that both relativistic as well as screening effects need to be considered in the calculation of two-photon ionization cross section. Relativistic effects significantly decrease the total cross section for heavy atoms, for the case of uranium, they decrease the cross section by a factor of two. Screening effects are highly sensitive to the photon energy and polarization as well as to the nuclear charge of the atom. In general, screening effects increase the cross section for low-$Z$ atoms by a factor of up to 1.5. However, for near-threshold photon energies, we observe a minimum in the total cross section which has pure screening origin. Due to a single allowed ionization channel, screening effects are most pronounced for ionization by circularly polarized light. For ionization of Ne, the cross section drops by a factor of three in the near-threshold energy region. Both, the relativistic as well as the screening effect will likely affect the photoelectron angular distribution of the two-photon ionization of neutral atoms. Therefore, 
it would be of great interest to use the theoretical formalism described above to also investigate the angular distribution, especially in the case, where the minimum of the cross section occurs. This will be the concern of our further study.\\

\begin{acknowledgments}
J.H. acknowledges the support from the Helmholtz Institute Jena. This work has been supported by the BMBF (Grant No. 05K13VHA).
\end{acknowledgments}

\end{document}